# Rapid Prediction of Three-Dimensional Scour Flow around Bridge Piers via Body-Fitted Coordinate-Based U-Net

Tokio Morimoto[1]†

1: Department of Civil Engineering, The University of Tokyo, Tokyo, Japan.

† Corresponding author

**Abstract (208 words)**

Predicting three-dimensional (3D) turbulent flows around bridge piers is a prerequisite for assessing local scour, a primary cause of infrastructure failure. While Computational Fluid Dynamics (CFD) captures complex flow features—such as horseshoe vortices—its high cost hinders real-time risk assessment. This study presents a physics-aware deep learning surrogate using a Body-Fitted Coordinate (BFC) system, BFC-UNet, designed to rapidly reconstruct 3D Reynolds-Averaged Navier–Stokes (RANS) solutions on curved domains. Unlike voxel-based Convolutional Neural Networks (CNNs) prone to staircase errors, the proposed architecture leverages a BFC system to predict the bed shear stress accurately. By transforming the physical O-grid into a canonical computational space, the model preserves the geometric integrity of the no-slip boundary. Trained on 2,304 simulations parameterized by inlet velocity and scour depth, BFC-UNet predicts velocity, pressure, and bed shear stress distributions with an $R^2$ value of > 0.98. It infers a full 3D domain ($2 \times 10^5$ cells) in just 8 milliseconds on a single Graphics Processing Unit (GPU)—achieving a speed-up of five orders of magnitude over the CFD solver. Crucially, the model captures the topological evolution of vortex structures, including wake expansion and diving flows. These findings position BFC-UNet as a promising foundation for real-time digital twins, bridging rigorous fluid mechanics with data-driven efficiency.

**List of symbols**

| | |
|---|---|
| $D$ | Pier diameter ( = 0.10 m) |
| $d$ | Scour depth |
| $l_1$ | Upstream scour width |
| $l_2$ | Downstream scour width |
| $p$ | Pressure |
| $Q$ | Second invariant of the velocity gradient tensor |
| $R^2$ | Coefficient of determination |
| $Re$ | Reynolds number |
| $U_{in}$ | Inlet velocity magnitude |
| $u$ | Streamwise velocity |
| $v$ | Spanwise velocity |
| $w$ | Vertical velocity |
| $w_i$ | Linear weighting coefficient |
| $x$ | Coordinate (streamwise direction) |
| $y$ | Coordinate (spanwise direction) |
| $z$ | Coordinate (vertical direction) |
| $\tau$ | Bed shear stress |

# 1. Introduction

The stability of hydraulic infrastructure is increasingly threatened by extreme rainfall events, which cause local scouring and eventually trigger bridge pier failure (Melville and Coleman, 2000; Raudkivi and Ettema, 1983). However, the flow field around a bridge pier is characterized by complex three-dimensional (3D) turbulent flow as shown in Figure 1. This flow field is dominated by interacting vortex systems (Baker, 1979; Dargahi, 1990; Roulund et al., 2005; Kirkil et al., 2008), notably the horseshoe vortex (upstream scouring), downflow (leading edge), and wake vortices (sediment entrainment). Consequently, the shear stress field on the riverbed surface, which drives the local scour, exhibits significant spatial variability. Furthermore, sediment transport and deposition—which dictate the evolution of bed height—are governed by the complex momentum balance between particles and fluid flow. This flow structure evolves dynamically as the local scour progresses, continuously interacting with the changing riverbed topography. Therefore, the scour phenomenon cannot be predicted by simple empirical correlations or global parameters; it demands resolving the underlying three-dimensional turbulent flow structures. Consequently, the ability to capture these local flow features is a prerequisite for reliable scour risk assessment.

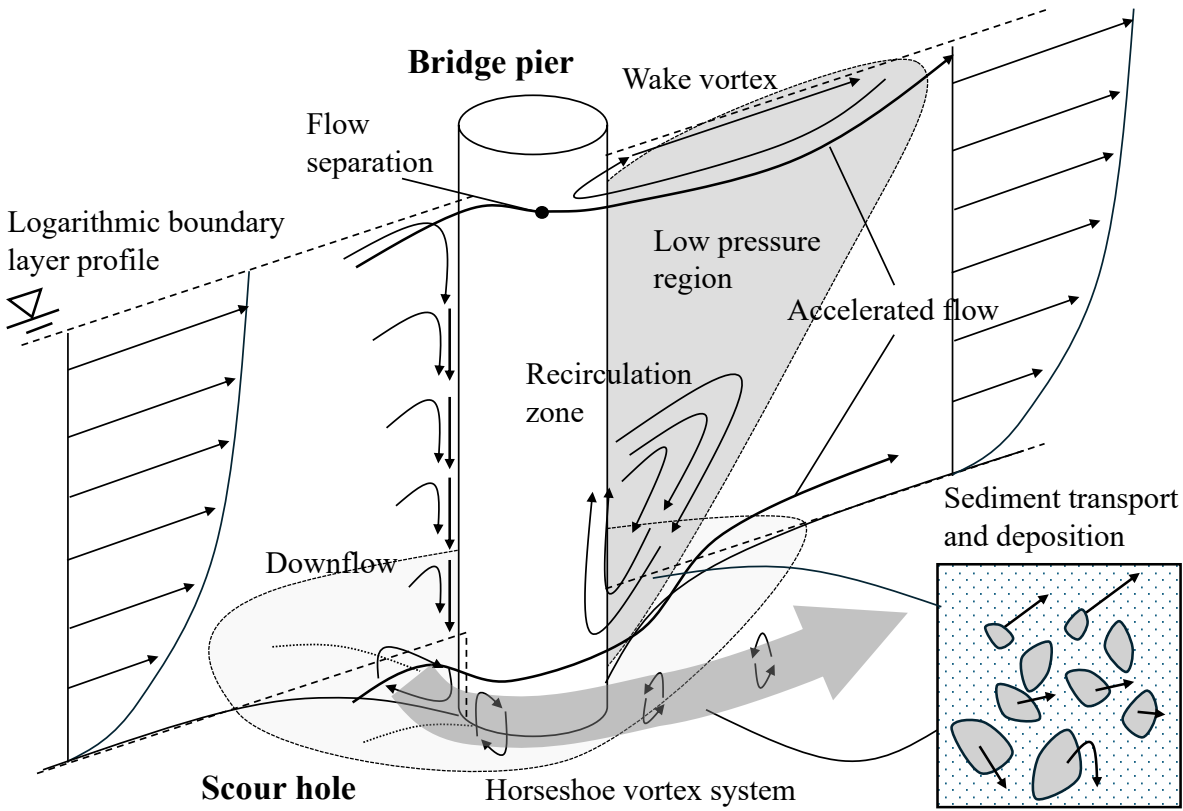

Figure 1. 3D turbulent flow around a bridge pier causing scour hole evolution and sediment transport.

Despite the critical need for detailed flow analysis, simulating the temporal evolution of scour morphology presents a fundamental time-scale discrepancy. The scouring process evolves over hours or days, whereas fluid dynamics requires resolving turbulent fluctuations on the order of fractions of a second. While Detached Eddy Simulation (DES) and Large Eddy Simulation (LES) can accurately capture the temporal fluctuations of vortices around a pier (Kirkil et al., 2008; Escauriaza and Sotiropoulos, 2011), these numerical methods are prohibitively expensive for simulating an entire local scour evolution. Most numerical approaches coupling Computational Fluid Dynamics (CFD) with sediment transport models, including the seminal work by Roulund et al. (2005), as well as recent solvers like SedFoam (Chauchat et al., 2017) and ibScourFoam (Xu and Liu, 2021; Song et al., 2022), employ the Reynolds-Averaged Navier-Stokes (RANS) equations to achieve high fidelity. However, their computational cost remains a significant barrier—simulating a single scour event often consumes days or weeks on high-performance computing (HPC) clusters. This bottleneck prohibits their use in real-time forecasting or parametric design studies requiring thousands of iterations, creating an urgent need for a computationally efficient surrogate model.

To overcome these limitations, data-driven surrogate models have emerged as a promising alternative. However, in the context of local scour, existing machine learning models have predominantly focused on predicting scalar quantities—such as the equilibrium scour depth—directly from global parameters (Firat and Gungor, 2009; Najafzadeh, M. and Azamathulla, 2013; He et al., 2023; Azma et al., 2024; Chimauriya et al., 2025; Ouallali and Taleb, 2025). While practically useful, these "black-box" approaches bypass the resolution of the flow field itself, failing to capture the governing fluid-structure interaction mechanisms.

The reconstruction of flow fields using machine learning has demonstrated growing potential (Wang et al., 2017; Fukami et al., 2019; Brunton et al., 2020; Vinuesa and Brunton, 2022). While surrogate modeling has been successfully applied to 2D flow fields to resolve flow physics (Bhatnagar et al., 2019; Thuerey et al., 2020; Maulik et al., 2021), extending these methods to 3D turbulent flows involving complex boundary conditions presents significant challenges. Standard approaches often struggle to balance geometric fidelity with computational efficiency when discretizing such three-dimensional domains. Voxel-based approaches, common in computer vision, are easily adaptable to Convolutional Neural Networks (CNNs) (Guo et al., 2016; Lienen et al., 2024) but fail to resolve the logarithmic boundary layer profile due to their "staircase" representation of curved surfaces. This lack of near-wall resolution leads to poor prediction of shear stress, the most critical parameter for scour evolution. Graph Neural Networks (GNNs) offer geometric flexibility and have been applied to mesh-based predictions (Sanchez-Gonzalez et al., 2020; Pfaff et al., 2021); however, their inference latency typically remains on the order of seconds (Suk et al., 2021; Zhao et al., 2024), which is still too slow for dynamic simulations requiring tens of thousands of time steps with a physical time increment less than 0.1 second (Roulund et al., 2005). Furthermore, Physics-Informed Neural Networks (PINNs) (Raissi et al., 2019; Eivazi et al., 2022), while theoretically attractive, currently incur excessive training costs for high-Reynolds-number 3D turbulence. Thus, there is a distinct lack of a surrogate model that simultaneously achieves the high spatial resolution required for boundary layer flows and the inference speed necessary for dynamic morphological evolution.

To address this critical gap, we propose a novel deep learning framework integrating classical Body-Fitted Coordinates (BFC) (Thompson et al., 1974) with a matured 3D U-Net architecture (Ronneberger et al., 2015; Çiçek et al., 2016), termed BFC-UNet. Unlike voxel-based methods that sacrifice geometric fidelity, or unstructured GNNs that compromise computational efficiency, our approach leverages the topological regularity of the structured O-Grid. We utilize Hermite interpolation to generate meshes that suppress non-orthogonality at the boundaries, ensuring that the physical domain—including the critical near-wall region—is mapped via a direct, lossless index-mapping into a canonical cubic latent space. Since the CFD mesh topology is structurally identical to the input tensor of the CNN, no voxelization or spatial interpolation is required during data pre-processing. This ensures that the high-resolution boundary layer information inherent in the original CFD results is preserved intact within the latent space. This structured data is then processed by a 3D U-Net, which is highly optimized for such regular grid structures. By applying a physics-aware weighting function to emphasize the no-slip boundaries, the model learns to reconstruct the 3D turbulent flow field, including the velocity components and pressure, with high fidelity.

The proposed model was trained on a dataset generated by OpenFOAM simulations, reducing the inference time to approximately 8 milliseconds—over five orders of magnitude faster than conventional CFD—while maintaining a high coefficient of determination for both flow velocity (mean relative $L_2$ error < 2.5%) and shear stress distribution (mean relative $L_2$ error < 7.5%). This paper demonstrates that the organic synergy between physics-aware grid generation and deep learning can provide an engineering-grade tool for the rapid prediction of dynamic scour evolution.

## 2. Methodology

### 2.1. Dataset Generation via CFD

To train the surrogate model, a comprehensive dataset of 3D turbulent flow fields around a bridge pier was generated using Computational Fluid Dynamics (CFD). The computational domain mimics a typical experimental flume setup, featuring a circular pier of diameter $D = 0.1$ m. The flow conditions were parameterized based on the Reynolds number ($Re$), varying between 20,000 and 60,000, consistent with the foundational experiments by Roulund et al. (2005).

The geometry of the scour hole was modeled parametrically to cover a wide range of erosion stages. We employed a randomized sampling strategy for the key geometric parameters, which are indicated in Figure 2:

- Inlet velocity ($U_{in}$): $0.2 - 0.6$ m/s
- Maximum scour depth ($d$): $0.01D - 0.6D$ m
- Upstream scour width ($l_1$): $1.0D - 1.5D$ m
- Downstream scour width ($l_2$): $1.5D - 2.5D$ m

#### 2.1.1. Mesh Generation: The Body-Fitted Coordinate (BFC) System

For high-fidelity CFD simulations of turbulent flow and accurate wall shear stress prediction, resolving the boundary layer is essential (Schlichting and Gersten, 2016). The Body-Fitted Coordinate (BFC) system is ideal for this purpose, as it produces a high-quality mesh that conforms to the object surface, maintaining orthogonality within the boundary layer (Thompson et al., 1974;

Sorenson, 1980). Rather than constructing a strict BFC system by solving Poisson's equation—which is computationally expensive—we employed an algebraic grid generation approach (Gordon and Hall, 1973; Smith, 1982). This explicit method is significantly faster, enabling the generation of over 2,000 training cases and facilitating future applications to dynamic local scour simulations where rapid re-meshing is required.

The computational domain is discretized using a structured O-grid topology, as shown in Figure 2. We utilized Hermite interpolation to construct the grid lines; this technique facilitates explicit control over the grid trajectory derivatives. By adjusting these derivatives, we effectively mitigated mesh skewness in the near-wall regions while ensuring topological robustness. This trade-off was essential to prevent mesh folding or negative volumes, particularly when generating grids for deep and steep scour holes where strict orthogonality constraints might compromise stability. To accurately capture the steep velocity gradients, the grid spacing in the wall-normal direction follows a geometric expansion; extremely fine cells are deployed adjacent to the no-slip boundaries to resolve the boundary layer ($y^+ < 1$), which then gradually coarsen towards the far field. This approach generates a body-fitted mesh that naturally captures the steep velocity gradients while maintaining topological stability to prevent mesh folding, even under complex scour geometries. The mesh resolution was fixed at $64 \times 64 \times 49$ (radial × azimuthal × vertical) cells, totaling approximately 200,000 cells.

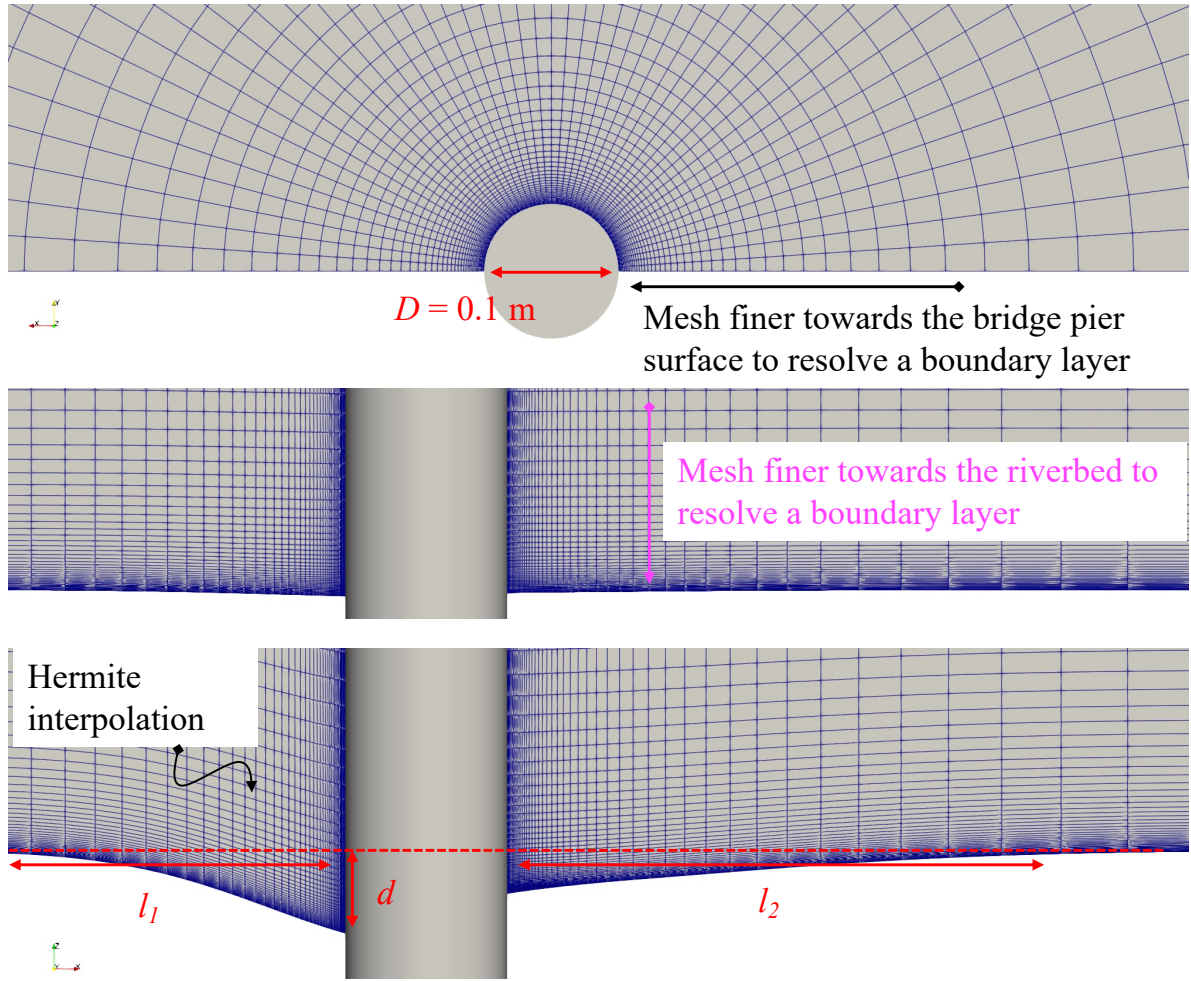

**Figure 2.** Schematic illustration of the structured O-grid topology and the parametric definition of the scour hole geometry. Key geometric variables ($D, d, l_1, l_2$) are randomized to cover various erosion stages.

### 2.1.2. Numerical Solver Settings

The Reynolds-Averaged Navier-Stokes (RANS) equations were solved using the $k\text{-}\omega$ SST turbulence model (Menter, 2003) within the OpenFOAM environment (Weller et al., 1998). Standard no-slip conditions were applied to the pier and riverbed, while a symmetry condition was imposed at the water surface (Figure 3). The simulations were iterated for 3,000 steps to reach a statistically steady state, ensuring continuity errors were reduced to $O(10^{-8})$. To compensate for any minor non-orthogonality in the grid arising from the complex geometry, two non-orthogonal correctors were employed to ensure numerical stability and accuracy. A total of 2,304 distinct cases were simulated on the Wisteria/BDEC-01 Odyssey supercomputer (Information Technology Center, The University of Tokyo), utilizing 576 cores (12 nodes) for approximately 12 hours.

## 2.2. BFC-UNet Architecture

### 2.2.1. Lossless Transformation: Mapping Physics to Tensor Space

The core innovation of the proposed "BFC-UNet" is the direct utilization of the structured grid topology. Unlike voxel-based methods that approximate geometry with staircase artifacts, our method maps the physical O-grid (Curvilinear coordinates) directly into a canonical cubic latent space (Cartesian coordinates) via a topologically isomorphic. We term this process a "lossless transformation" as it introduces no interpolation errors during the pre-processing stage (Figure 3).

The network input consists of 4 channels: the coordinate values ($x, y, z$) and the inlet velocity ($U_{in}$). Standard Convolutional Neural Networks (CNNs) rely on translational invariance—applying the same kernels across the entire domain. However, turbulent flows exhibit strong spatial dependency, particularly near boundaries where the "Law of the Wall" governs, which is fundamentally different from the free-stream behavior. To bridge this gap, we explicitly treat the grid coordinates as input channels. This explicit spatial encoding acts as a positional embedding, enabling the network to recognize the proximity to the no-slip boundaries and effectively capture the steep velocity gradients within the thin boundary layer, which standard CNNs often fail to resolve. Explicitly including the coordinates allows the network to implicitly learn the metric tensor (Jacobian) of the coordinate transformation, thereby understanding the physical distortion of the cells.

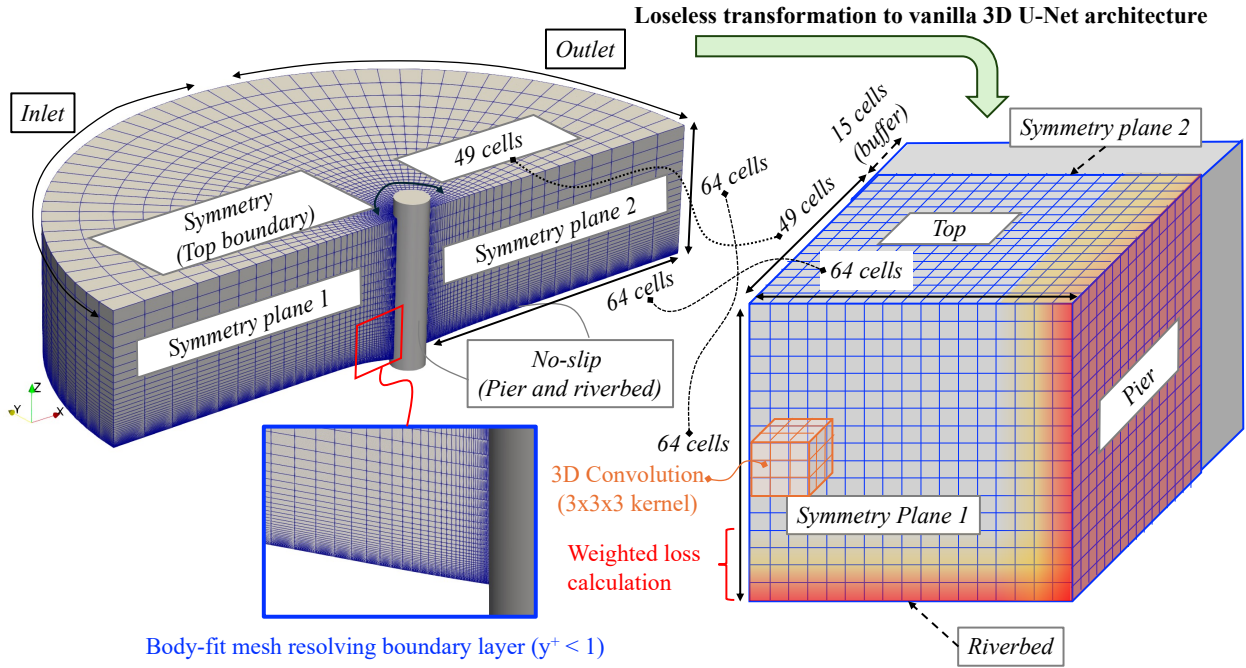

**Figure 3. Rendered CFD simulation domain with boundary conditions and its "loseless transformation" to 3D-UNet architecture.**

### 2.2.2. Network Structure

The architecture is based on a standard 3D U-Net as illustrated. in Figure 4. It features an encoder-decoder structure with skip connections to preserve high-frequency spatial information.

- **Convolution:** $3 \times 3 \times 3$ kernels with ReLU activation.
- **Pooling/Upsampling:** Max pooling for downsampling and trilinear interpolation for upsampling.

- **Physics-Aware Symmetry Padding:** The computational domain represents a symmetric half of the physical setup. To strictly satisfy the symmetry boundary condition at the domain edges (the azimuthal boundaries), we utilized a padding strategy that replicates the boundary values into the buffer region. This opration creates a zero-gradient condition across the boundary, effectively preventing unphysical flow across the symmetry plane and minimizing boundary artifacts during the convolution operations.

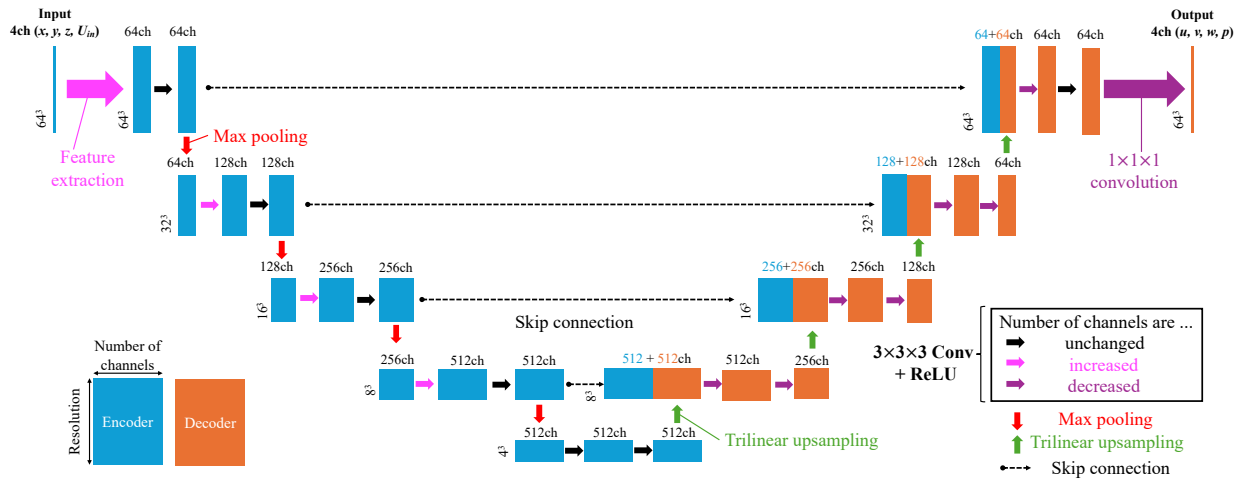

**Figure 4. Schematic diagram of 3D UNet architecture predicting the velocity and pressure fields from the grid coordinates and the inlet velocity.**

## 2.3. Training and Inference

### 2.3.1. Weighted Loss Function

A standard Mean Squared Error (MSE) loss is insufficient for this problem because the majority of the computational domain consists of the free stream, where flow gradients are small. Without intervention, the network tends to prioritize minimizing errors in these volumetric regions, neglecting the thin but physically critical boundary layer. To address this imbalance, we designed a spatially weighted loss function. The loss $L$ is defined as:

$$L = \frac{1}{N}\sum_{i=1}^{N} w_i \|\boldsymbol{u}_{CFD}^{i} - \boldsymbol{u}_{UNet}^{i}\|^2 + \|p_{CFD}^{i} - p_{UNet}^{i}\|^2$$

where $w_i$ is a linear weighting coefficient that increases from 1 (freestream) to 10 (wall boundaries). This forces the network to prioritize the accurate reconstruction of the boundary layer profiles, which are critical for accurate bed shear stress prediction.

### 2.3.2. Training Process

The network was implemented in PyTorch. The dataset was split into 80% for training and 20% for validation. The model was trained for 300 epochs using the Adam optimizer (Kingma and Ba, 2015) with an initial learning rate of $1 \times 10^{-3}$. Figure 5 (The learning curve) illustrates the convergence history. The validation loss decreases monotonically alongside the training loss, indicating that the model generalizes well to unseen geometries without overfitting. The validation loss reached a plateau ($< 0.012$) around 298 epochs. The inference speed of the trained BFC-UNet is approximately 8 milliseconds per case on a single Graphics Processing Unit (GPU) (NVIDIA A100), representing a speed-up factor of $10^5$ compared to the conventional CFD solver.

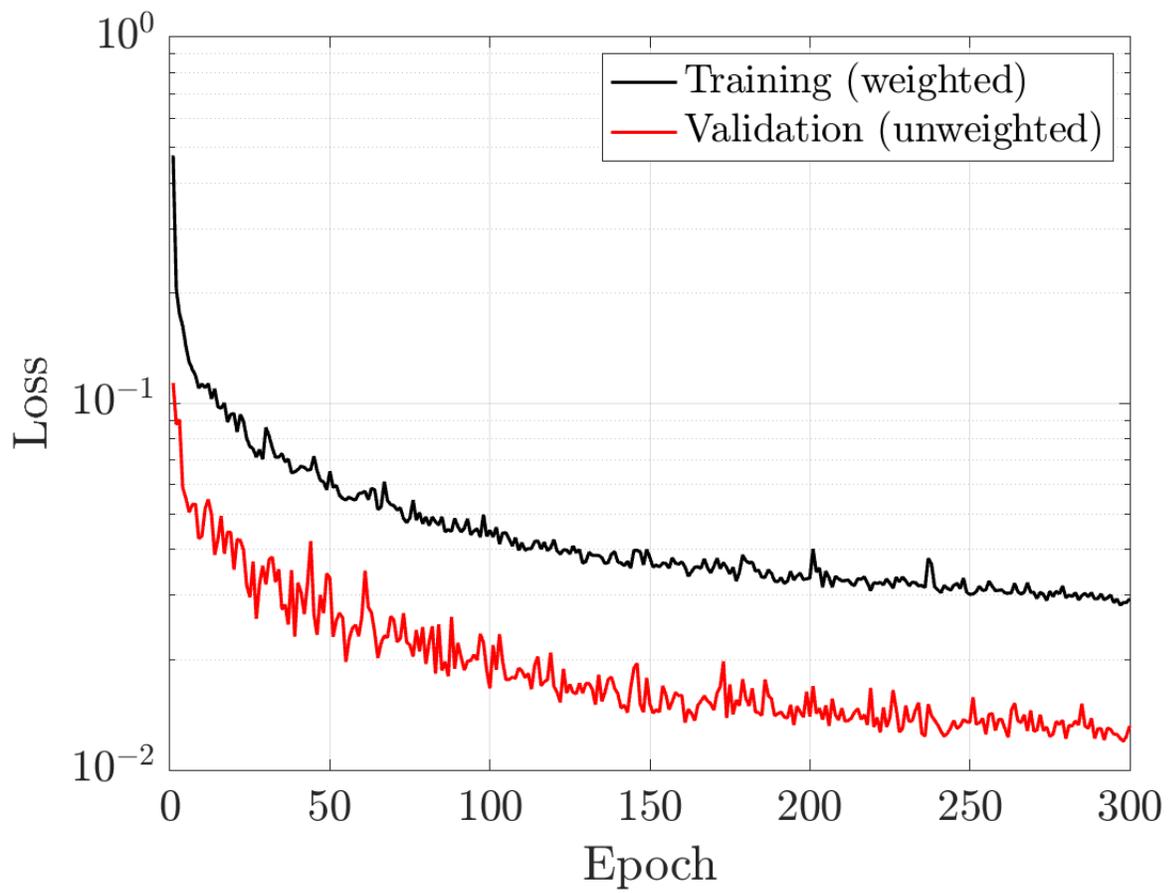

Figure 5. Training and validation curves.

## 3. Results

### 3.1. Quantitative Accuracy and Robustness

First, we evaluate the global error distribution across all test cases to assess the statistical robustness of the model. Figure 6 illustrates the distribution of prediction errors with respect to the scour depth ratio ($d/D$), where markers are colored based on the inlet velocity magnitude. As shown in the scatter plot, the model demonstrates high fidelity across the entire parameter space: 92.6% of the test cases exhibit a global velocity error of less than 5%. Furthermore, the mean error of the near-wall velocity gradient—a critical metric for shear stress estimation—remains below 10% for the majority of cases (86.9%). Crucially, no significant correlation is observed between the error magnitudes and either the scour depth or the inlet velocity. This indicates that the BFC-UNet is robust against the geometrical changes associated with scour progression and is not biased toward specific flow regimes.

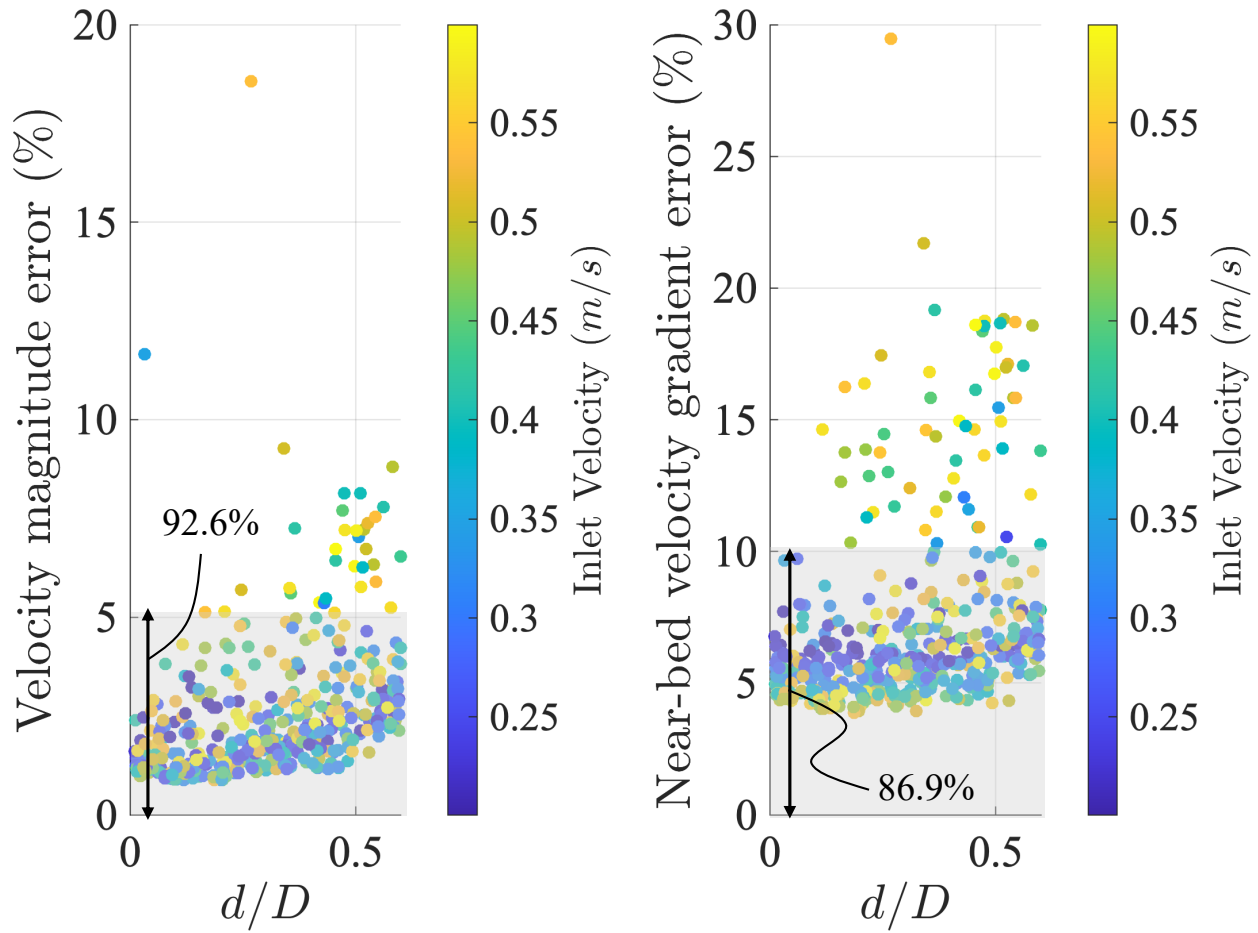

**Figure 6. Relative L$_2$ errors of velocity magnitude and near-bed velocity gradient (bed shear stress proxy) against the scour depth ratio $d/D$**

### 3.2. Quantitative Validation via One-Dimensional Profiles

To further validate the local accuracy of the BFC-UNet, we examine the one-dimensional profiles of velocity components ($u,w$) and pressure ($p$) along the domain centerline ($y=0$).

### 3.2.1. Streamwise Variations and Wake Characteristics

Figure 7 compares the streamwise distributions of the flow variables at different heights ($z$) for three representative scour depths ($d/D$ = 0.04, 0.27, 0.58). The results demonstrate excellent agreement between the CFD ground truth (solid lines) and the BFC-UNet predictions (circles) across all scour stages.

- Streamwise Velocity ($u$): The model accurately captures the sharp flow deceleration approaching the pier leading edge ($x/D \approx -0.5$). In the wake region ($x/D > 0.5$), the model correctly predicts the negative velocities and the recovery profile. Notably, the BFC-UNet successfully reproduces the shortening of the recirculation zone length as the scour depth increases ($d/D = 0.04 \rightarrow 0.58$), a phenomenon driven by the expansion of the flow passage within the scour hole.
- Vertical Velocity ($w$): The vertical velocity profiles reveal a dynamic shift in the vortex structure. As shown in the middle row of Figure 7, the flow structure evolves from a predominantly horizontal shear flow at low scour depths to a structure with significant vertical momentum transfer at $d/D = 0.58$. The model nearly perfectly predicts this transition, including the strong downflow at the pier front—which is the primary driver for scour evolution—and the complex upwelling in the wake.
- Pressure ($p$): The pressure distributions (bottom row) confirm that the driving forces are also well-resolved. The model captures both the pressure build-up at the stagnation point ($x/D = -0.5$) and the subsequent pressure recovery downstream, ensuring that the pressure gradients governing sediment transport are physically consistent.

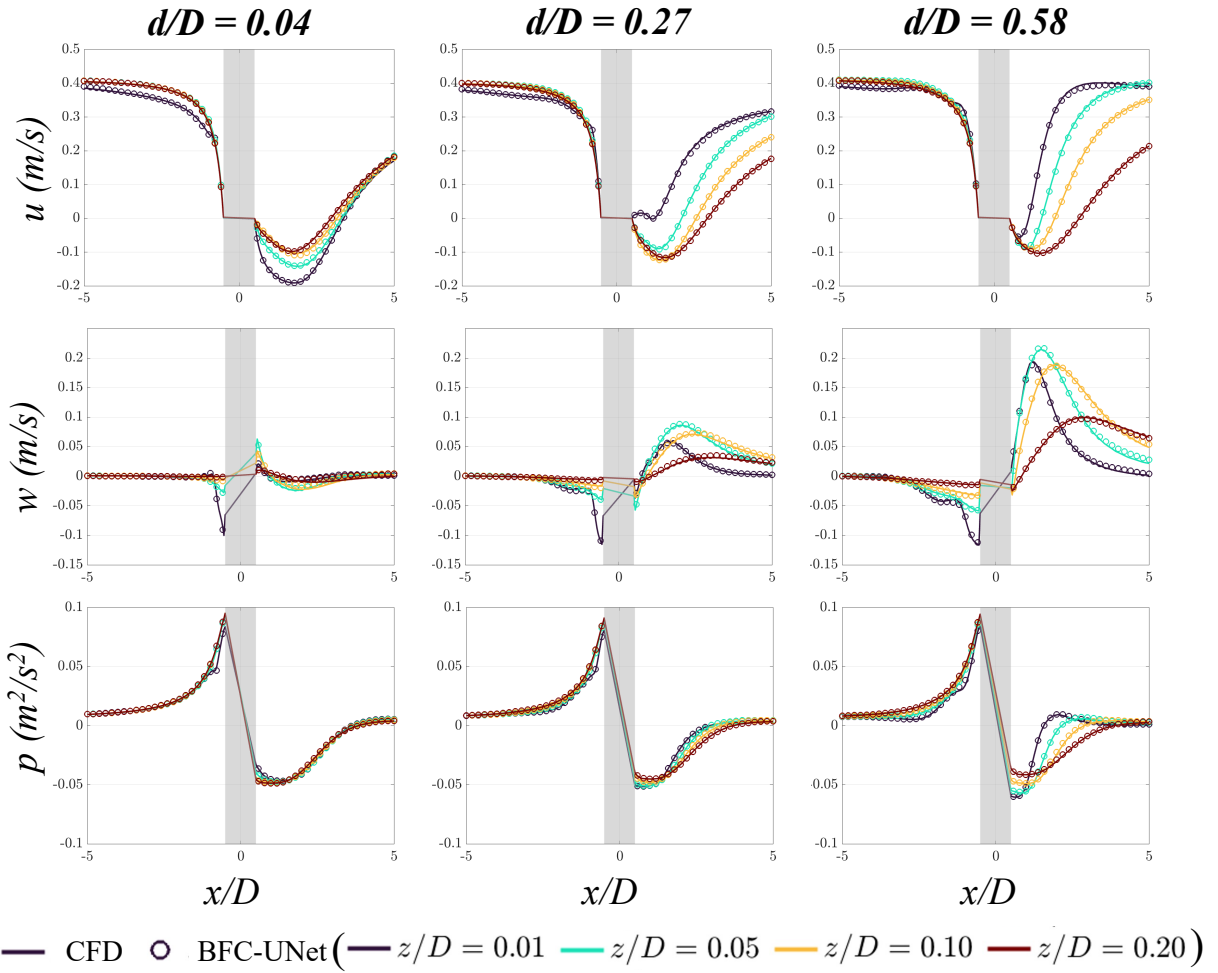

Figure 7. Comparison of streamwise profiles of velocity components ($u$, $w$) and pressure ($p$) along the centerline ($y = 0$). The columns correspond to different scour depths: $d/D = 0.04$, $0.27$, and $0.58$. The shaded gray area indicates the cylinder position.

### 3.2.2. Vertical Profiles and Boundary Layer Resolution

Resolving the vertical flow structure, particularly within the scour hole and the near-wall boundary layer, is essential for accurate sediment transport modeling. Figure 8 presents the vertical profiles of the flow variables at selected streamwise locations upstream ($x/D < 0$) and downstream ($x/D > 0$) of the pier.

- Streamwise Velocity ($u$): The model demonstrates high fidelity in capturing the boundary layer characteristics. The logarithmic nature of the approaching flow and the complex shear layers in the wake are both well-preserved. Notably, the BFC-UNet accurately predicts the evolution of the horseshoe vortex system at the upstream pier face ($x/D = -0.55$), where the region of reverse flow near the bed expands significantly as the scour depth increases from $d/D = 0.04$ to $0.58$.
- Vertical Velocity ($w$): The vertical velocity profiles highlight the model's ability to capture topological flow changes. For the initial flat bed case ($d/D = 0.04$), w remains negligible across most of the domain. However, as the scour progresses to $d/D = 0.58$, complex downflow and upflow structures emerge due to the intensification of the horseshoe vortex and the wake recirculation. The BFC-UNet successfully predicts this transition from a 2D-like flow to a fully 3D flow structure within the scour hole.
- Pressure ($p$): The vertical pressure distribution also exhibits significant changes. For low $d/D$ values, while the pressure varies significantly in the streamwise direction (high pressure at the stagnation point and low pressure in the wake), its vertical profile remains largely uniform. In contrast, the deep scour hole ($d/D = 0.58$) induces complex three-dimensional flow structures, leading to significant vertical pressure gradients near the bed.

The proposed model successfully captures these complex vertical variations, which are critical for determining the vertical force balance on sediment particles.

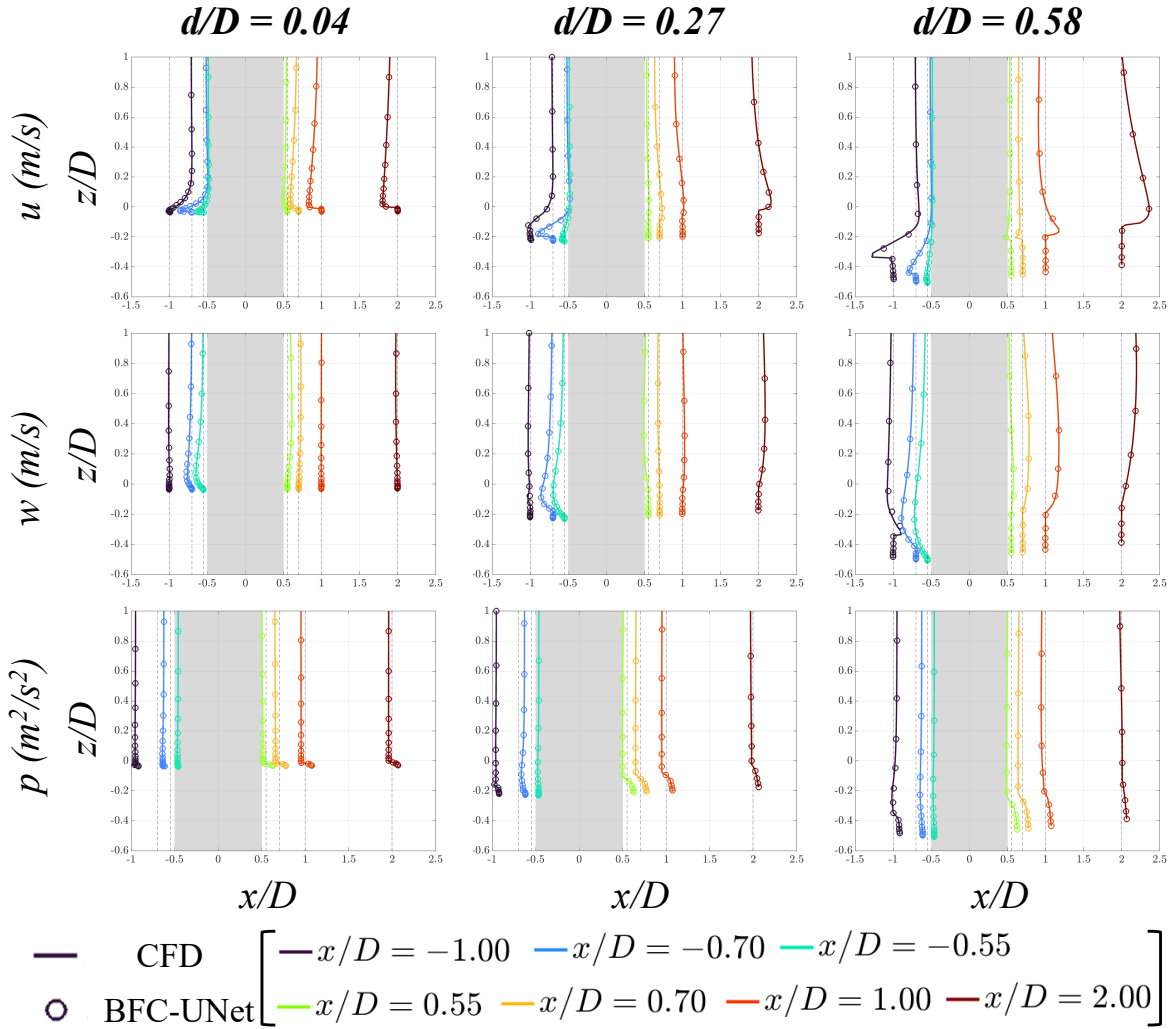

**Figure 8. Vertical profiles of mean flow variables at various streamwise locations ($x/D$). The rows correspond to streamwise velocity $u$ (top), vertical velocity $w$ (middle), and pressure $p$ (bottom). The columns correspond to different scour depths: $d/D = 0.04, 0.27$, and $0.58$. For visualization purposes, the profiles are offset by their streamwise location: the plotted quantity is $\phi + x/D$, where $\phi$ represents the respective variable ($u$, $w$, or $p$) normalized by**

the inflow parameters. The vertical dashed lines indicate the zero-reference for each profile at the corresponding $x/D$ station.

## 3.3. Overall Statistical Performance

Finally, Table 1 summarizes the quantitative performance of the BFC-UNet across the entire test dataset, evaluated using the mean relative $L_2$ error and the coefficient of determination ($R^2$).

The model achieves excellent global accuracy, with relative errors of 1.97% for $u$, 2.75% for $p$, and 2.45% for the velocity magnitude $||u||$. The near-wall velocity gradient (shear stress proxy) is also predicted with a low error of 7.09%, confirming the model's applicability to sediment transport problems. Table 1 also reports the $R^2$ values of the local profile assessment shown in Figures. 7 and 8 for a variety of the scour depths. The scores for $u$ and $p$ exceed 0.99 across all scour depths. While the $R^2$ for w is slightly lower for the initial stage ($d/D = 0.04$, $R^2 = 0.86$), this is attributed to the negligible magnitude of w in the flat-bed case (where the signal-to-noise ratio is low) —as observed in the corresponding panels of Figures. 7 and 8—rather than a model deficiency. For advanced scour stages where vertical motion is significant ($d/D = 0.58$), the $R^2$ for w improves to 0.9788, demonstrating the model's capability to resolve complex 3D flow features when they are physically present.

**Table 1. Quantitative summary of the BFC-UNet performance for a variety of flow variables.**

| Variables | Primary variables | | | | |
|---|---|---|---|---|---|
| | Global performance | | Local profile assessment ($R^2$) | | |
| | Whole domain | | Validation lines in Figs 7 and 8 | | |
| | Mean rel. $L_2$ error (%) | Average $R^2$ | $d/D = 0.04$ | 0.27 | 0.58 |
| Streamwise vel. $u$ | 1.97% | 0.9985 | 0.9990 | 0.9982 | 0.9963 |
| Vertical vel. $w$ | 10.5% | 0.9834 | 0.8597 | 0.9733 | 0.9788 |
| Pressure $p$ | 2.75% | 0.9986 | 0.9900 | 0.9974 | 0.9947 |
| Secondary and derived variables | | | | | |
| Velocity mag. $\|\boldsymbol{u}\|$ | 2.45% | 0.9984 | – | – | – |
| Spanwise vel. $v$ | 3.51% | 0.9977 | – | – | – |
| Wall shear stress proxy | 7.09% | 0.9929 | – | – | – |

### 3.4. Qualitative Results: Flow Field Reconstruction

To visually assess the model's capability in resolving complex turbulent structures, we compare the CFD-predicted fields with the BFC-UNet reconstructions across representative scour stages.

### 3.4.1. Symmetry Plane Velocity and Flow Topology

Figure 9 illustrates the streamwise velocity distribution on the central symmetry plane ($y = 0$) for scour depths of $d/D$ = 0.04, 0.27, and 0.58. As the scour hole deepens, the flow topology undergoes significant changes: the high-momentum downflow at the leading edge extends deeper into the scour hole, and the recirculation zone behind the pier expands. The BFC-UNet successfully reproduces these topological evolutions with high fidelity. Notably, the expansion of the wake region and the flow acceleration along the scoured bed are accurately captured. The difference plots (bottom row) show a predominantly neutral color, confirming that the error is minimal throughout the bulk flow region, even as the domain geometry deforms significantly. Localized discrepancies are observed near the upstream pier face; however, this is primarily attributed to the "double penalty2 phenomenon arising from a slight spatial shift in the sharp gradient of the horseshoe vortex core, rather than a failure to predict the structure itself.

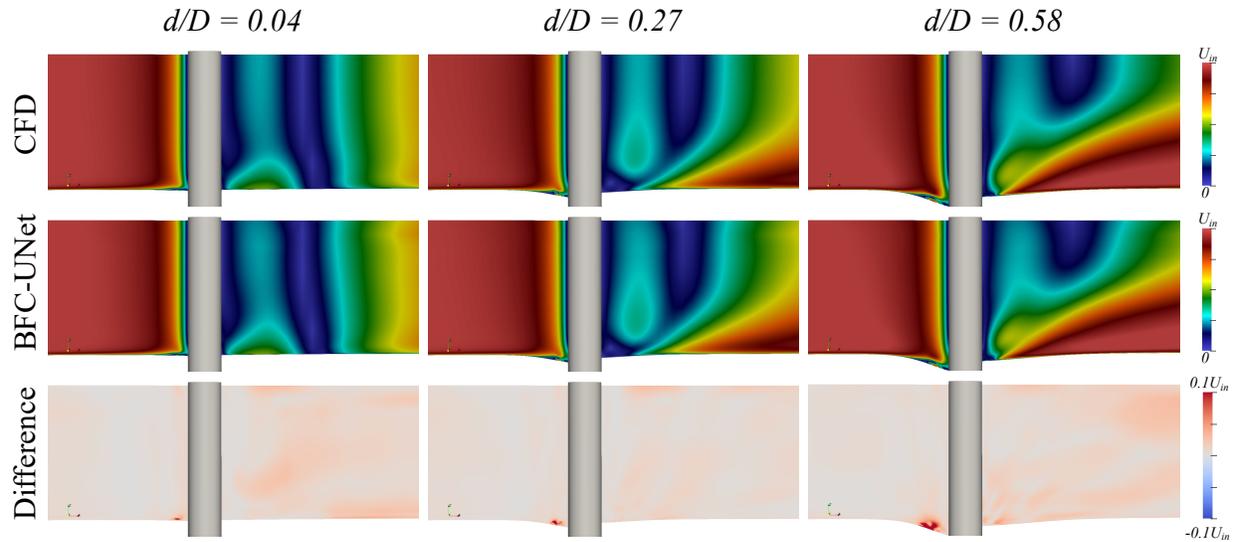

**Figure 9.** Symmetry plane velocity field obtained from the CFD (Top) and the BFC-UNet (Middle) for different scour depths ($d/D$ = 0.04, 0.27, 0.58). The difference of the fields was visualized at the bottom.

### 3.4.2. Pressure Field Distribution

The prediction of the pressure field is critical for evaluating the fluid forces acting on the bridge pier. Figure 10 compares the pressure distributions on the symmetry plane. The model accurately reconstructs the high-pressure stagnation zone at the pier's leading edge and the low-pressure wake region downstream. Crucially, the vertical pressure gradient within the scour hole—which drives the secondary flows—is well-preserved. The difference plots indicate that the discrepancies are negligible.

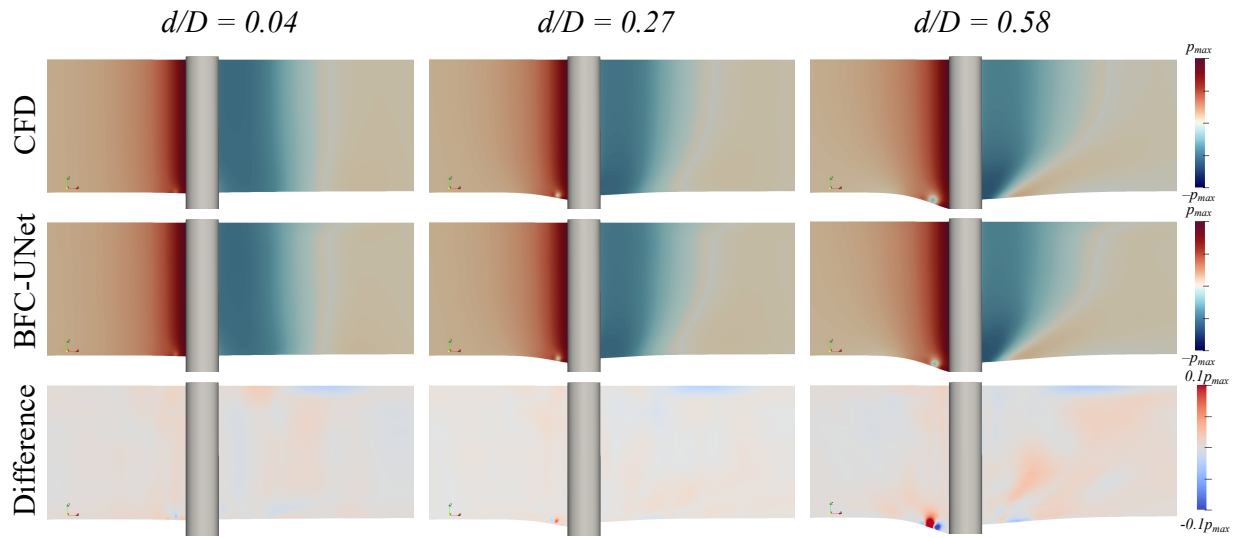

**Figure 10. Symmetry plane pressure field obtained from the CFD (Top) and the BFC-UNet (Middle) for different scour depths ($d/D$ = 0.04, 0.27, 0.58). The difference of the fields was visualized at the bottom.**

### 3.4.3. Bed Shear Stress

Figure 11 presents the bed shear stress distributions, which are the primary drivers of sediment transport. The CFD results (top row) show the characteristic amplification zones at the sides of the pier, which expand as the scour hole deepens ($d/D$ = 0.04 → 0.58) due to flow contraction. The BFC-UNet (middle row) captures this expansion pattern nearly perfectly. The difference plots (bottom row) are mostly featureless, indicating that the model correctly predicts not only the magnitude but also the spatial extent of the high-stress regions. This capability is vital for coupling the BFC-UNet with morphological models, where the bed shear stress drives the bed evolution.

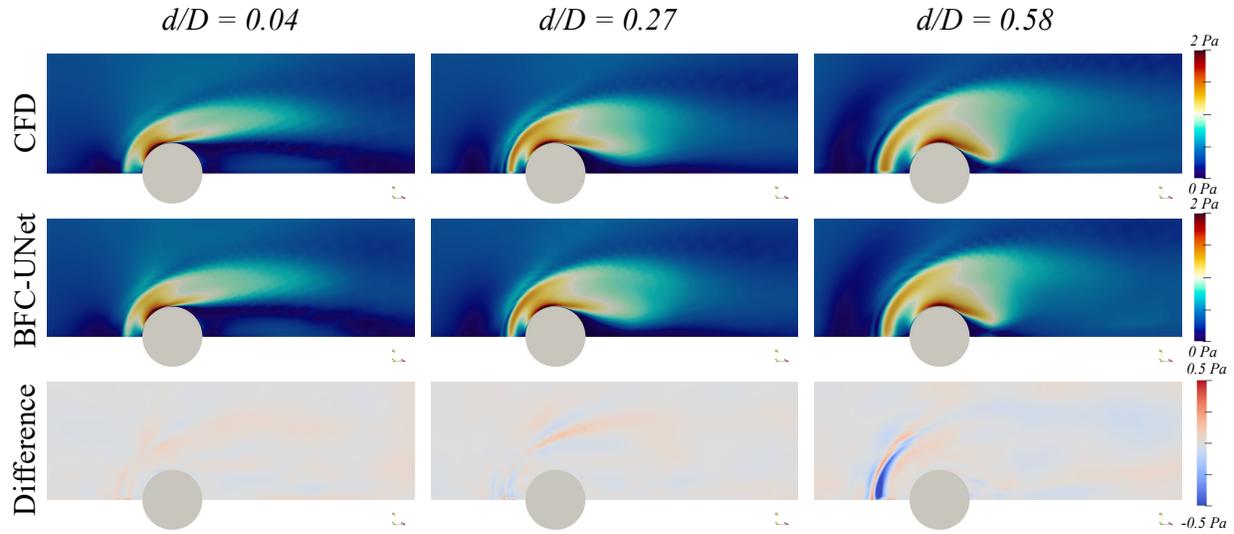

**Figure 11.** Bed shear stress distribution obtained from the CFD (Top) and the BFC-UNet (Middle) for different scour depths ($d/D$ = 0.04, 0.27, 0.58). The difference in the distributions was visualized at the bottom.

### 3.4.4. Three-Dimensional Coherent Structures

To verify the resolution of coherent flow structures, we visualize the vortex systems using the Q-criterion isosurfaces ($Q = 25s^{-2}$) colored by velocity magnitude (Figure 12). The complex interplay between the horseshoe vortex system at the pier base and the recirculation wake is clearly visible. The BFC-UNet (right panel) reproduces these topological features with remarkable detail. The necklace-like vortex wrapping around the upstream base (horseshoe vortex system) is clearly identified. The arch-shaped vortices rising toward the free surface in the wake are well-reconstructed. The ability to predict these coherent structures indicates that the BFC-UNet has learned the underlying physics of vorticity transport, rather than merely memorizing pixel-wise statistical correlations.

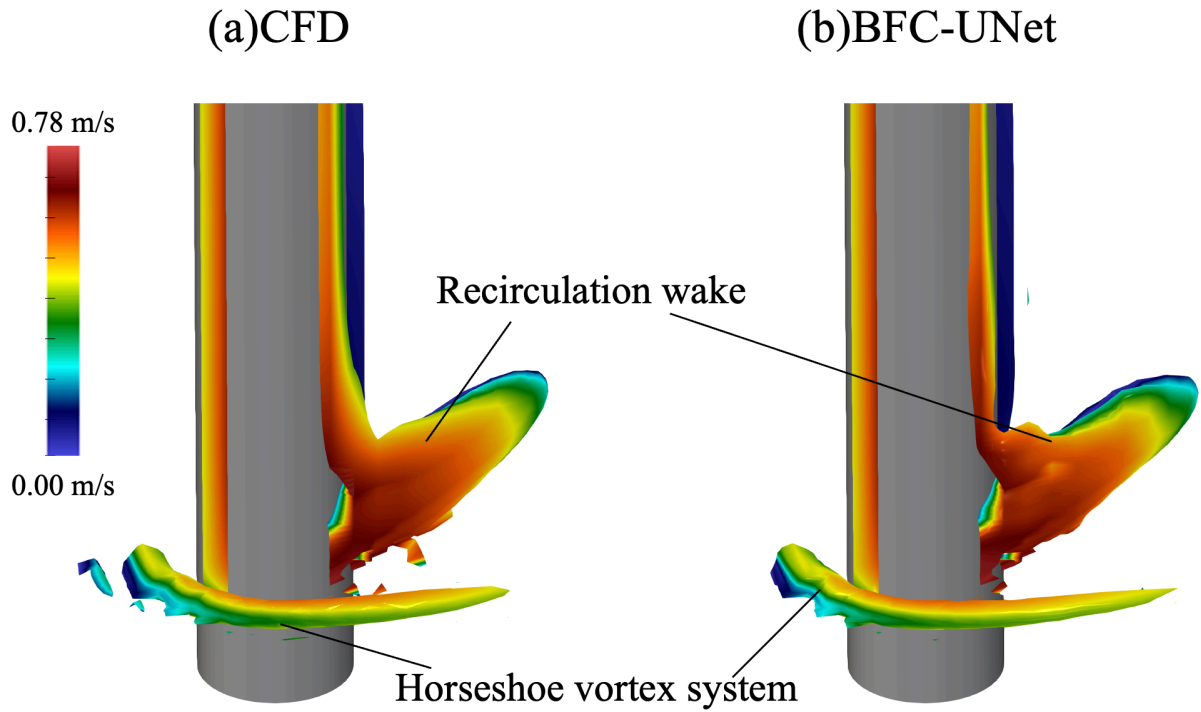

**Figure 12: Comparison of instantaneous three-dimensional vortex structures visualized by $Q$-criterion iso-surfaces ($Q = 25$ s$^{-2}$). The surfaces are colored by velocity magnitude. (a) CFD benchmark. (b) BFC-UNet prediction.**

### 3.4.5. Lagrangian Particle Trajectories

Finally, to visualize the transport pathways relevant to sediment motion, Figure 13 presents the 3D streamlines generated by the BFC-UNet for two distinct scour stages ($d/D$ =0.04 and 0.58). Streamlines are released from identical upstream locations near the bed ($z/D < 0.05$).

At the initial stage ((a) $d/D = 0.04$), the flow remains largely horizontal (quasi-two-dimensional), bypassing the cylinder with minimal vertical deflection. In contrast, the presence of the deep scour hole ((b) $d/D = 0.58$) induces a strong diving flow into the pit, followed by a complex helical upward motion in the wake. This "corkscrew" motion is responsible for lifting sediment particles from the scour hole into the water column. The BFC-UNet successfully captures this intricate, topography-induced flow transition, demonstrating that the model preserves the continuity and rotationality of the flow field.

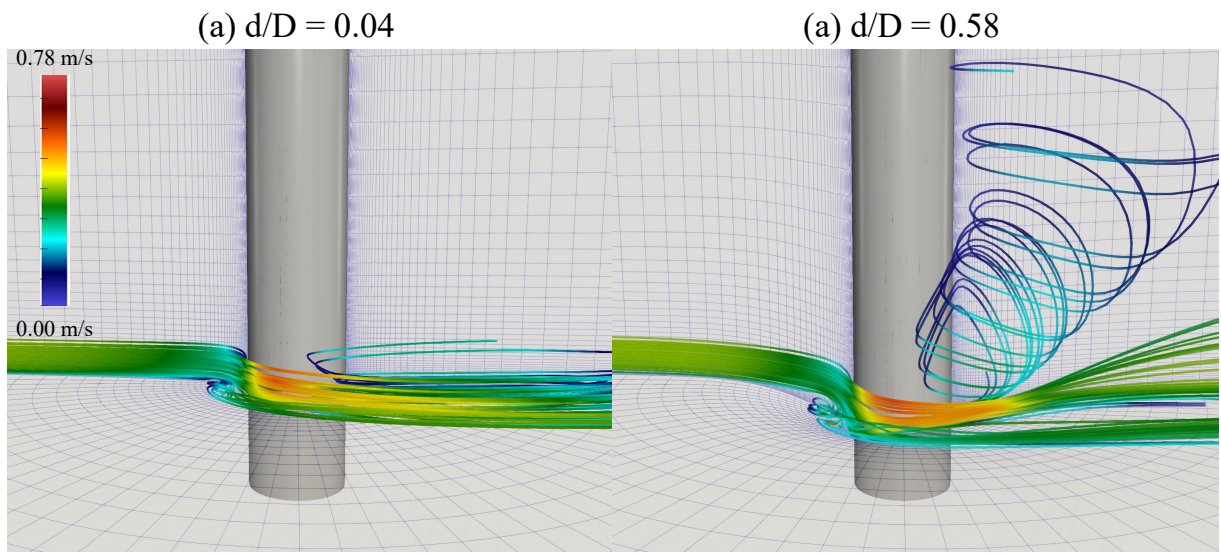

**Figure 13: Comparison of three-dimensional streamlines predicted by the BFC-UNet model for different scour depths. (a) $d/D = 0.04$ and (b) $d/D = 0.58$.**

## 4. Conclusions

This study addressed the challenge of predicting complex three-dimensional turbulent flows around a bridge pier with a developing scour hole, a problem that has traditionally demanded prohibitive computational resources. We successfully developed a data-driven surrogate model, BFC-UNet, which achieves an unprecedented balance between fluid dynamic fidelity and computational efficiency.

The distinct advantage of the proposed method lies in its rigorous preprocessing strategy rooted in domain knowledge. By employing a Body-Fitted Coordinate (BFC) system governed by boundary layer theory and ensuring orthogonality via Hermite interpolation, we effectively transformed the geometrically complex physical domain into a canonical cubic computational space. This topologically consistent mapping ("lossless transformation") allowed us to utilize a standard ("Vanilla") 3D U-Net architecture without the need for complex geometric deep learning techniques or unstructured data handling. The neural network could thus focus purely on learning the physics of the flow—such as the horseshoe vortex system and wake separation—without struggling against geometric irregularities or interpolation errors.

The performance of the BFC-UNet is quantitatively and qualitatively remarkable. The model reproduced the detailed 3D flow structures with a coefficient of determination exceeding 0.98 for developed scour holes. Crucially, this high fidelity was achieved with an inference time of approximately 8 milliseconds on a single GPU, representing a speed-up of over $10^5$ times compared to conventional CFD solvers. The entire workflow, from generating 2,304 training cases using OpenFOAM (12 hours) to training the model (4 hours), was completed within a single day. These results demonstrate that the computational cost is no longer a bottleneck for 3D scour analysis.

In conclusion, this study proves that high-performance surrogate modeling does not necessarily require increasingly complex neural architectures. Instead, it highlights the critical importance of physics-aware data representation. The success of BFC-UNet stems from the organic synergy between the classical wisdom of grid generation and the pattern-recognition power of deep learning.

The ultra-fast inference capability of BFC-UNet opens new horizons for hydraulic engineering. Since the topological structure of the O-grid remains consistent as long as the bed slope satisfies the angle of repose, this model can be seamlessly integrated into a time-stepping loop for dynamic scour simulation. This would enable real-time prediction of scour hole evolution, a task previously deemed impossible. Furthermore, the framework is readily extensible to different pier geometries—similar to airfoil parametrization in aerodynamics—allowing for the exploration of optimal pier shapes that inherently mitigate scour risk.

**Acknowledgements**

This work was supported by JSPS KAKENHI Grants Number 24K17346. This research was conducted using the FUJITSU Supercomputer PRIMEHPC FX1000 and FUJITSU Server PRIMERGY GX2570 (Wisteria/ BDEC-01) at the Information Technology Center, The University of Tokyo.

**Author declaration: Conflict of Interest**

The authors have no conflicts to disclose.

**Data availability**

Data will be made available on request.

**Declaration of generative AI and AI-assisted technologies in the manuscript preparation process**

During the preparation of this work the author used Gemini Pro 3.0 in order to improve the readability and language of the manuscript. After using this tool/service, the author reviewed and edited the content as needed and take full responsibility for the content of the published article.